\documentclass{article}
\usepackage{graphicx} 
\usepackage[noblocks]{authblk}
\usepackage{amssymb}
\usepackage{indentfirst}
\usepackage{amsmath}
\usepackage{mathptmx}
\usepackage{graphicx}
\usepackage{comment}
\usepackage{xspace}
\usepackage{tabularx}
\usepackage{float}
\usepackage{longtable}
\usepackage{algorithm}
\usepackage{algorithmic}
\usepackage{mathtools}
\usepackage{amsmath}
\usepackage{authblk}
\usepackage{hyperref}
\usepackage{tabularx}
\usepackage{caption}
\usepackage{subcaption}
\usepackage{booktabs}
\usepackage{multirow}
\usepackage{cite}
\usepackage{adjustbox}
\usepackage{pifont}
\usepackage{url}
\usepackage{placeins}
\usepackage[a4paper, margin=1in]{geometry}

\title{Optimizing Resource Allocation to Mitigate the Risk of Disruptive Events in Homeland Security and Emergency Management}

\author[1]{Parastoo Akbari}
\author[1]{Cameron A. MacKenzie}
\affil[1]{Department of Industrial and Manufacturing Systems Engineering, Iowa State University, Ames, IA, USA}

\begin{document}
\date{}
\maketitle
\begin{abstract}
\normalsize
Homeland security in the United States faces a daunting task due to the multiple threats and hazards that can occur. Natural disasters, human-caused incidents such as terrorist attacks, and technological failures can result in significant damage, fatalities, injuries, and economic losses. 
The increasing frequency and severity of disruptive events in the United States 
highlight the urgent need for effectively allocating resources in homeland security and emergency preparedness. 
This article presents an optimization-based decision support model to help homeland security policymakers identify and select projects that best mitigate the risk of threats and hazards while satisfying a budget constraint. The model incorporates multiple hazards, probabilistic risk assessments, and multidimensional consequences and integrates historical data and publicly available sources to evaluate and select the most effective risk mitigation projects and optimize resource allocation across various disaster scenarios. We apply this model to the state of Iowa, considering 16 hazards, six types of consequences, and 52 mitigation projects. Our results demonstrate how different budget levels influence project selection, emphasizing cost-effective solutions that maximize risk reduction. Sensitivity analysis examines the robustness of project selection under varying effectiveness assumptions and consequence estimations. The findings offer critical insights for policymakers in homeland security and emergency management and provide a basis for more efficient resource allocation and improved disaster resilience.
\end{abstract}

\section{Introduction}
The United States is vulnerable to different disruptive events including natural, human-caused such as terrorist attacks, and technological disasters, each of which can lead to fatalities and injuries, infrastructure damage, and significant financial losses.
Billion-dollar weather and climate disasters have been increasing in both frequency and severity in the United States \cite{noaa2025}. The United States experienced at least 28 and 27 billion-dollar weather events in 2023 and 2024, respectively. The annual cost from these billion-dollar weather events has averaged almost \$150 billion from 2019-2024. A probabilistic time-series model calculates a 10\% chance the total cost of billion-dollar disasters in the United States could exceed \$175 billion and 1\% chance the total cost could surpass \$500 billion \cite{shukla2024}. Most of the U.S. federal assistance for immediate disaster response is funded through the Disaster Relief Fund (DRF). From 2013-2019, total DRF obligations ranged from \$9 to \$27 billion. From 2020-2024, total DRF obligations ranged from \$37 to \$79 billion \cite{painter2024}.
Protecting against cyber threats has become a top priority for homeland security whether the threat originates from individual criminals or nation-state actors \cite{albert2023homeland}.  

A key function of homeland security is preparing for emergencies and disasters. Emergency management consists of four phases: mitigation, preparedness, response, and recovery \cite{spiewak2015}. Emergency preparedness seeks to mitigate the risk from natural and human-caused hazards and enhance a country or community's ability to respond to a severe event. Emergency preparedness remains a significant challenge due to the significant variability in the appearance and evolution of these disruptive events, insufficient investment in emergency preparedness at all levels of government, challenges of coordinating emergency preparedness among multiple organizations, and a lack of citizen preparedness to these events. For example, the Los Angeles Fire Department response to the Los Angeles wildfires in 2025 seems to have been hampered by budget cuts and a failure to take measures that could have mitigated the severity of these wildfires \cite{ingram2025,pringle2025,depuykamp2025}.

Quantifying risk frequently means assessing the probability and consequences of risk \cite{Kaplan1981, shokrgozar2025,yaghmaei2022temporal}. Translating the output of risk assessments into meaningful decision support and risk management is especially challenging in the homeland security \cite{Haimes2011}. In the United States, the Department of Homeland Security (DHS) and state and local governments identify projects and activities that mitigate the risk for a wide variety of threats and hazards. Several reasons make identifying the best set of projects and activities to pursue a difficult public policy decision. First, disruptions are uncertain and knowing the exact timing, geographic area, and damage of a disruption is impossible. Some of this uncertainty derives because threats may be driven by criminals, terrorists, and nation-states that seek to damage the nation. Second, the consequences of disruptions are multidimensional \cite{Medeiros2021} and may include fatalities and injuries, infrastructure damage, business losses, and environmental impacts \cite{Lundberg2018, RussellLundberg2015, suresh2024value}. Third, there are competing priorities for resources at the federal, state, and local level. Allocating resources in many organizations is a complex problem. It is hardly ever done effectively, and it is frequently a satisficing effort \cite{HerbertASimon1979}. Resource allocation in emergency preparedness and homeland security does not follow a strict risk-optimization framework that can achieve the best possible risk reduction. 

There is an opportunity to allocate resources more effectively to create a better return on investment in emergency preparedness and homeland security. This article designs a decision support model that accounts for the previously discussed factors (e.g., uncertain disasters, multidimensional consequences) in order to better represent the complexity of preparing for mutliple hazards. We propose an optimization model that can enable DHS and state and local governments to select a set of risk mitigation projects while adhering to budget constraints. Risk is represented as the probability of an event and the multidimensional consequences from the occurrence of the event. This model is unique in its ability to simultaneously account for the probability of several disruptive events, their multidimensional consequences (e.g., fatalities, property damage, customers without power), differences in the effectiveness of projects to mitigate risk, and the trade-offs inherent in resource allocation. By incorporating these factors into a single optimization framework, our model seeks to address the gaps in current practices and provide a more practical approach to risk mitigation.

The remainder of this paper is organized as follows. Section 2 provides a detailed literature review, examining the existing research and identifying gaps in resource allocation for homeland security. Section 3 outlines the optimization model, including the objective function, decision variables (i.e., risk mitigation projects), and how the decision variables impact the objective function. Section 4 applies the resource allocation model to the state of Iowa's mitigation decisions that includes 16 hazards, 6 types of consequences, and 53 potential risk mitigation projects. The probabilities and consequences for each hazard and the effectiveness and cost of each risk mitigation project are calculated and estimated from a plethora of information sources and publicly available databases. We solve the resource allocation model for different budgets in order to generate insights from the model on the most cost-effective risk mitigation projects. Section 5 concludes the paper by summarizing insights, discussing limitations, and proposing directions for future research.

\section{Literature Review}
DHS was established after the September 11 terrorist attacks to protect America from terrorism and other threats and hazards. Operations research has contributed to addressing homeland security problems by providing methodologies to optimize decision making and resource allocation \cite{Wright2006}. McLay \cite{McLay2015} conducted a comprehensive survey of discrete optimization models in homeland security across the four different phases of emergency management (mitigation, preparedness, response, and recovery). Optimizing resource allocation in homeland security seeks to reduce the risks and vulnerabilities before disruptions occur and to improve response and recovery efforts when events occur. Deterministic and stochastic optimization models are frequently suggested as tools to allocate resources for a variety of homeland security problems. These models have been applied in areas such as aviation security \cite{Nie2008, Xiaofeng2012}, port security and screening \cite{Orosz2009, RebeccaDreiding2010}, critical infrastructure protection \cite{Norkin2018, AlMannai2008}, the resilience of the electric power grid \cite{MacKenzie2016}, terrorism protection \cite{Atkinson2010}, fire protection \cite{Raj2017, Behrendt2019}, and emergency medical services \cite{EricDuBois2020, Xiang2016, Larson2013, Altay2013}. 

Some studies focus on how DHS should allocate money for homeland security projects at the state and local community levels. Research \cite{willis2007, Hu2011} examines the extent to which the Urban Area Security Initiative, which allocates money to cities, aligns with risk of terorrism that each city faces. Greenberg et al. \cite{Greenberg2009} use publicly available data to optimize the allocation of homeland security funds to states with the goal of protecting electricity-generating capacity. Their model incorporates political equity and population size as well as risk-related criteria. Resources can be allocated before a disruption occurs to help states prevent and prepare for disruptive events \cite{MacKenzie2021} or they can be allocated after a disruption to help a region or different industries recover \cite{MacKenzie2016}. 

Allocating resources in homeland security and emergency management should balance among the risks, costs, and benefits of making those allocations \cite{Abkowitz2012, Lin2012}. 
Farrow \cite{FARROW2007} introduces six foundational models for evaluating homeland security expenditures. The models address issues such as technological limitations, behavioral interactions, and decision making under uncertainty.
Winterfeldt et al. \cite{vonWinterfeldt2020} present a methodology for risk-informed benefit-cost analysis of homeland security research products and test their proposed methodology on 10 U.S. Coast Guard projects.
A probabilistic, risk-based framework for prioritizing among homeland security research and development problems evaluates the future value with uncertainty in both the threat environment and technology development \cite{Stromgren2018}. Kim et al. \cite{Kim2022} use deep learning methods to improve the predictability of natural disasters and uses cost-benefit analysis to quantify the cost effectiveness of mitigation projects.

Game theory has emerged as a popular tool for analyzing and addressing homeland security problems where terrorists and other individuals intend to attack the United States. Game theory approaches can provide a robust framework for modeling and analyzing interactions between defenders and adversaries. These studies consider challenges such as how to optimally allocate a budget when defending against multiple threats to the homeland \cite{BrownG2008, JingZhang2018, Nikoofal2012, Aziz2020, Shan2013, Zhang2018, Zhang2019, Zhuang2011, Wang2016, Long2024, Guan2016, Wei2024, King2016, Norkin2018, Golany2012, Ge2019, FALLAH2010}.

This article contributes to the homeland security risk assessment literature by identifying several hazards---including severe weather, criminal or terrorist attacks, and pandemics---based on the the Fedeal Emergency Management Agency's (FEMA) Threat and Hazard Identification and Risk Assessment (THIRA)\cite{HSEMD2021}. Most studies in the literature consider a single or perhaps a few different hazards. We improve on existing methodologies by calculating probabilities based on historical data and incorporating multidimensional consequences (e.g., fatalities, injuries, infrastructure damage, property damage, business losses) of each disruptive event. The core contribution is a mathematical optimization model to help a decision maker choose projects to minimize the risk (probability and consequences) from these hazards subject to a budget constraint. The projects can prevent, mitigate, and/or protect against a hazard.  
The model serves as a comprehensive decision-support tool for optimizing resource allocation in risk management.

\section{Mathematical Model}
The resource allocation problem can be formulated as a mathematical model in which a decision maker is choosing a set of projects in order to mitigate the risk of several disruptive events. The total cost of these projects is subject to a budget constraint. This article defines risk as a function of two factors: the probability of a disruptive event and its consequences. A total of $n$ disruptive events exist, each occurring with a probability $\hat{p}_i$, $i=1,2,\ldots,n$ assuming no mitigation activities. These events are not mutually exclusive or collectively exhaustive, i.e., $\sum_{i=1}^n \hat{p}_i$ could be less than or greater than 1. A total of $m$ types of consequences exist. Given a disruptive event $i$, consequence $j$, $j=1,2,\ldots,m$ is denoted as $\hat{f}_{ij}$ assuming no mitigation activities. 

Each project $k$ that could be selected to mitigate the risk of these $n$ disruptive events is represented by a decision variable $x_k$, and $x_k=\left\{0,1\right\}$ (either the project is selected for funding or not), where $k=1,2,\ldots,K$ and $K$ is the total number of possible projects. Vector $\mathbf{x}$ represents all of the mitigation projects under consideration. Some mitigation projects can reduce the probability of the disruptive event; some mitigation projects can reduce the consequences from the disruptive event; and some mitigation projects can reduce both the probability and consequences. Some mitigation projects can only reduce the risk of a single event and other mitigation projects can reduce the risk of multiple events. 

The occurrence probability, $p_i$, of disruptive event $i$, is a function of the project allocation decisions $\mathbf{x}$ and is denoted as:

\begin{equation}\label{eq:prob_event}
p_i\left(\mathbf{x}\right) = \hat{p}_i \prod_{k=1}^K \alpha_{ik}^{x_k}
\end{equation}

\noindent where $0\leq \alpha_{ik} \leq 1$ is the effectiveness of project $k$ in reducing the probability of event $i$. Smaller values of $\alpha_{ik}$ indicate greater effectiveness of project $k$ in reducing the probability of event $i$. The parameter $\alpha_{ik}$ could be equal to 1 for many events and projects, which indicates a project $k$ does not reduce the probability of event $i$.

Similarly, consequence $j$ resulting from disruptive event $i$ is a function of the allocated mitigation projects $\mathbf{x}$, and it is denoted by $f_{ij} \left(\mathbf{x}\right)$. Each function $f_{ij} \left(\mathbf{x}\right)$ can be defined similarly to $p_i\left(\mathbf{x}\right)$:

\begin{equation}\label{eq:cons_event}
f_{ij}\left(\mathbf{x}\right) = \hat{f}_{ij} \prod_{k=1}^K \beta_{ijk}^{x_k}
\end{equation}

\noindent where $0\leq \beta_{ijk} \leq 1$ describes the effectiveness of project $k$ in reducing the consequence $j$ for event $i$. Smaller values of $\beta_{ijk}$ indicate greater effectiveness for project $k$. If $\beta_{ijk}=1$, then project $k$ has no impact on consequence $j$ if event $i$ occurs. 

Since the disruptive events can result in multiple types of consequences, the resource allocation problem can be modeled as multi-objective decision, or the consequences can be combined into a single objective function. This article takes the latter approach by combining the consequences into a single objective through an additive function $\sum_{j=1}^m w_j f_{ij}\left(\mathbf{x}\right)$ for each event $i$ where $w_j$ represents a constant marginal trade-off between consequence $j$ and the other consequences. In this research, $w_j$ represents a dollar value associated with each consequence (e.g., value of a fatality, value of an injury).

We assume the decision maker wants to allocate projects in order to minimize the expected consequences, as given in the following optimization problem:

\begin{equation}\label{eq:optimization}
\begin{aligned}
\underset{\mathbf{x}}{\text{minimize}} \quad & \sum_{i=1}^n \hat{p}_i \prod_{k=1}^K \alpha_{ik}^{x_k} \left( \sum_{j=1}^m w_j \hat{f}_{ij} \prod_{k=1}^K \beta_{ijk}^{x_k} \right) \\
\textrm{subject to} \quad & \mathbf{c}^\intercal \mathbf{x} \leq B \\
& x_k \in \left\{0,1\right\}, \; \; \; k=1,2,\ldots, K
\end{aligned}
\end{equation}

\noindent where $\mathbf{c}$ is a vector containing the cost of each mitigation project and $B$ is the total allocation budget.

\section{Application}
The resource allocation model is applied to the state of Iowa in order to select projects to mitigate the risk of several disruptive events. The decision maker is the Iowa Department of Homeland Security and Emergency Management (HSEMD), who is responsible for preparing the state of Iowa for emergencies and disasters. The resource allocation model aims to support HSEMD decision makers in prioritizing and selecting risk mitigation projects subject to a budget constraint. State departments of homeland security identify threats and hazards in order to determine capabilities required for emergency preparedness and response. The THIRA is the starting point for emergency planning at the state level \cite{HSEMD2021}.
Parameter estimation for the resource allocation model derives from publicly available data, government documents, journal articles, and news stories. 

\subsection{Assumptions and parameter estimation}
Iowa HSEMD identifies $n=16$ hazards that threaten the state: animal disease, train derailment, flood, cyberattack, drought, earthquake, extreme heat, wildfire, hazardous materials (hazmat) release, human disease, improvised explosive device (IED) attack, tornado, radiological incident, dam failure, winter storm, and a bridge failure \cite{HSEMD2021,HSEMD2023}. Seven of these hazards are extreme weather or geological events; two are diseases; two would result from malicious actors; and five hazards are technological or infrastructure failures (which could be accidents or intentionally caused by individuals).
These hazards could result in scores of different impacts. We select $m=6$ different types of consequences to include in the resource allocation model. The consequences are fatalities, injuries, property damage, crop damage, closed businesses, and customers without power. These six consequences seem to broadly cover the range of impacts that decision makers in Iowa care about and are impacts for which we could find data. 

Table \ref{tab:hazards} depicts the probabilities and consequences for each hazard if no projects are selected to mitigate the risk. These parameters were extracted and estimated from multiple sources. This subsection provides greater detail on how we estimated parameters for two hazards, a winter storm and a dam failure, in order to illustrate how parameters were estimated for the resource allocation model. 

\begin{table}
    \centering
     \caption{Probability and consequences for each hazard assuming no risk mitigation}
    \label{tab:hazards}
     \begin{adjustbox}{max width=\textwidth}
    \begin{tabular}{llrrrrrr} 
        \hline
         \multirow{3}{*}{Hazard} & \multirow{3}{*}{Probability} & \multirow{3}{*}{Fatalities} & \multirow{3}{*}{Injuries} & \multirow{3}{*}{\begin{tabular}{c}Property\\  damage (\$) \end{tabular}} & \multirow{3}{*}{\begin{tabular}{c}Crop\\  damage (\$) \end{tabular}} & Customers  & \multirow{3}{*}{\begin{tabular}{c}Businesses\\  closed \end{tabular}} \\
         & &&&&& without &\\
         &&&&&& power &\\ \hline
         Animal disease & 0.00415 & 0  & 0  & 0  & 0  & 0  & 704 \\
         Train derailment & 0.01785 & 6.35 & 85.88  & 52,500,000  & 0  & 0 & 3\\
         Flood & 0.4545  & 0 & 20 & 177,710,000 & 249,361,667 & 738 & 136 \\
         Cyberattack & 0.0208 & 0 & 0 & 0 & 0 & 4749 & 0\\
         Drought & 0.2045 & 0 & 0 & 71,683,333 & 567,664,444 & 773 & 229\\
         Earthquake & 0.0075 & 30 & 3114 & 18,350,000,000 & 0 & 679 & 60\\
         Extreme heat & 0.1590 & 0.43 & 6.571 & 0 & 0 & 893 & 146\\
         Wildfire & 0.0909 & 0 & 1.25 & 278,750 & 10,000 & 50 & 0\\
         Hazmat release & 0.296 & 0 & 420 & 0 & 0 & 0 & 28 \\
         Human disease & 0.0099 & 8443 & 500,461 & 0 & 0 & 0 & 2800\\
         IED attack & 0.0238 & 10.6 & 54.36 & 51,250,000 & 0 & 0 & 12\\
         Tornado & 0.6363 & 1.178 & 40.857 & 48,436,927 & 6,628,975 & 853 & 81\\
         Radiological & \multirow{2}{*}{0.000001} & \multirow{2}{*}{0.4} & \multirow{2}{*}{0} & \multirow{2}{*}{637,560,000} & \multirow{2}{*}{0} & \multirow{2}{*}{0} & \multirow{2}{*}{120}\\ 
         incident &&&&&&&\\
         Dam failure & 0.04586 & 0 & 0 & 14,321,875 & 0 & 1462 & 227\\
         Winter storm & 0.2045 & 0.444 & 5.555 & 3,614,722 & 29,888,889 & 5000 & 271\\
         Bridge failure & 0.0209 & 14.16 & 21.33 & 562,500,000 & 0 & 0 & 0\\ \hline
    \end{tabular}
    \end{adjustbox}
\end{table}

The National Oceanic and Atmospheric Administration (NOAA) publishes the Storm Events Database \cite{NationalCentersforEnvironmentalInformationNCEI2024} consisting of severe weather events that occurred in each county in the United States since 1950. The database for Iowa classifies 48 distinct types of weather events which we categorized into the 6 natural hazards in Table \ref{tab:hazards}: flood, drought, extreme heat, wildfire, tornado, and winter storm. The Storm Events Database publishes the number of fatalities and injuries, property damage in dollars, and crop damage in dollars for these weather events. 

As part of the data preprocessing, we excluded events in the Storm Events Database \cite{NationalCentersforEnvironmentalInformationNCEI2024} prior to 1980. The data in the Storm Events Database are often incomplete due to limitations in data collection and processing technologies. Weather, climate, and the environment changes over time, and older data may be less suitable for assessing current conditions.
Rows with missing values in injuries, fatalities, property damage, and crop damage were removed. We identified and removed duplicate entries. 

As a state agency, HSEMD focuses on severe events. Winter storms occur multiple times in every year in Iowa. We define a severe winter storm as a winter storm in the Storm Events Database that meets at least one of the following citeria;
\begin{itemize}
    \item The number of injuries exceeds 5. 
    \item The number of deaths exceeds 5.
    \item At least \$100 million in property damage occurs.
    \item At least \$50 million in crop damage occurs.
\end{itemize}
Based on these criteria, 9 winter storm events occurred in Iowa over a span of 44 years. The annual probability of a winter storm is $9/44=0.2045$. The average number of fatalities and injuries and the average property and crop damage for these 9 events are included in the consequences for winter storm, as depicted in Table \ref{tab:hazards}.

The Storm Events Database 
does not provide information on infrastructure damage (e.g., power outages) or lost business. HSEMD estimates quasi-worst-case numbers for many impacts resulting from the 16 hazards, including fatalities, injuries, customers without power, and closed businesses \cite{HSEMD2021}. We used regression to model the relationship between the logarithm of fatalities and injuries according to HSEMD and the logarithm of fatalities and injuries that we assessed for the hazards (e.g., from the Storm Events Database). The regression model used the number of customers without power and the number of business closed for the winter storm scenario according to HSEMD as inputs to estimate those same parameters depicted in Table \ref{tab:hazards} for a severe winter storm in our resource allocation model. 


The likelihood and consequences for a dam failure in Iowa are derived from a couple of sources. The National Inventory of Dams \cite{NationalInventoryofDams2024} provides detailed information with 70 data fields for each dam in the United States. According to this database, 4,054 dams exist in Iowa out of a total of 91,807 dams nationwide, representing approximately 4.41\% of all dams in the United States. 

The Association of State Dam Safety Officials publishes the Dam Incident Data \cite{AssociationofStateDamSafetyOfficials2024}, which is the most comprehensive source of historical data on dam failure incidents in the United States. Since the National Dam Safety Program was established between 1978 and 1981, we considered incidents from 1980 to 2023. Incidents involving dams classified as having high potential hazard were included in the analysis. If an incident occurs, a dam with high potential hazard can lead to fatalities and significant economic and environmental damage. From 1980 to 2023, 46 high-hazard dam failure incidents occurred in the United States, resulting in an annual rate of 1.04 serious dam failures. 
To calculate the probability of an extreme dam failure event in Iowa, we multiplied the annual rate of dam failure by the proportion of dams in Iowa relative to the United States. The probability of a serious dam failure in Iowa is $1.04*0.0441=0.04586$.

None of these 46 high-hazard dam failure incidents resulted in fatalities or injures, but the average property damage was \$14,321,875. We found no data on crop damage from a dam failure, and we assume \$0 in crop damage. The number of customers without power and businesses closed are estimated similar to how those parameters are estimated for a winter storm. 

The parameters for the other 14 hazards were estimated using a similar approach as the winter storm and dam failure. For each hazard, we used relevant databases or research reports to filter and preprocess the data, identified incidents that meet severity criteria, and calculated probabilities and consequences. This ensured consistency in the estimation process across different hazards. 

Table \ref{tab:dollars} provides the dollar value for each consequence $w_j$. Property damage and crop damage are already measured in dollars and $w_j=1$ for those two consequences. The dollar values for fatalities and injuries were obtained from the National Risk Index, which assigns \$11.6 million per fatality and \$1.16 million per injury \cite{FEMA2022}. The dollar value for each customer without power and for each businesses closed were derived from \cite{LeoraLawton2003, Tierney1995} and adjusted to 2024 prices using the Consumer Price Index. 

\begin{table}
    \centering
    \caption{Dollar value per consequence}\label{tab:dollars}    
    \begin{tabular}{lr}
          \hline  
         Fatality & \$11,600,000\\
         Injury & \$1,160,000\\
         Customer without power & \$2,195.54\\
         Business closed & \$103,075.79\\
         \hline
    \end{tabular}
\end{table}

The potential risk mitigation projects (i.e., the decision variables) consist of 52 projects in the 2023 Iowa Comprehensive Emergency Plan \cite{HSEMD2023}. These projects and their associated costs are displayed in Table \ref{sec:appendixa} in the Appendix. Different methods were used to estimate the cost of each project. First, some projects---or very similar projects---appear with 
an associated cost in the 2018 Iowa Comprehensive Emergency Plan \cite{HSEMD2018}. Second, if a project did not appear in the 2018 Iowa Comprehensive Emergency Plan, we relied on two FEMA databases to estimate the cost of each project: Hazard Mitigation Assistance (HMA) Projects \cite{FEMA_HazardMitigation2025} and HMA Subapplications \cite{FEMA_HMA2025}. We classified each Iowa project according to the categorization scheme of those two databases. 
For example, Project 10, which states that jurisdictions acquire software or other tools to help with implementing codes or regulations that mitigate hazards, corresponds to the Codes and Standards category in the HMA Subapplications, which includes activities such as technical assistance and the development, administration, and enforcement of new codes. Since the FEMA databases contain multiple projects in the same category as the Iowa project, we averaged the costs of these FEMA projects to estimate the cost of the Iowa risk mitigation project.

\begin{figure}[H]
    \centering
    \includegraphics[width=\linewidth]{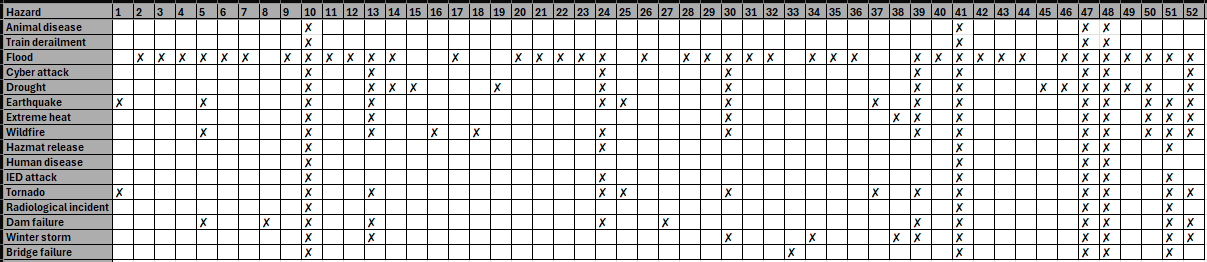}
    \caption{Projects and their associated hazards}
    \label{fig:project-hazard mapping}
\end{figure}

The effectiveness parameters for each mitigation project depend on our assessment of the hazards that each project can help mitigate. We use the project's description to determine if a project can mitigate a hazard. If a project does apply to a hazard, we also determine if the project would reduce the probability of the hazard and/or if it would mitigate none, some, or all six of the consequences. Figure \ref{fig:project-hazard mapping} depicts the hazards that each project helps to mitigate. If a project does not reduce a hazard's probability, then $\alpha_{ik}=1$ and if the project does not reduce a specific consequence, then $\beta_{ijk}=1$. 


The 2023 Iowa Comprehensive Emergency Plan assigned a priority letter grade, A, B, C, or D, to each of the 52 mitigation projects. This letter grade reflects, in part, the effectiveness of these projects. We translate these letter grades to the effectiveness parameters $\alpha_{ik}$ and $\beta_{ijk}$, as shown in Table \ref{tab:effectiveness}. We assume that a project is more effective at reducing the consequences than reducing the probability. Four projects (10, 41, 47, and 48) mitigate all 16 hazards, and we assume that these four projects are half as effective at reducing the consequences as the projects that are more targeted to some of the hazards. These four projects have general goals that apply to all hazards, and it seems reasonable to assume that they are less effective than a project with specific actions to mitigate one or possibly a few hazards. For example, Project 41 that mitigates all of the hazards receives a letter grade of A. Its effectiveness parameter $\beta_{ij,47}=0.9$ instead of $0.8$ as provided in Table \ref{tab:effectiveness}.
\vspace{1cm}
\begin{table}[H]
    \centering
    \caption{Values for effectiveness parameters}\label{tab:effectiveness}    
    \begin{tabular}{ccc}
          \hline  
          \multirow{2}{*}{Letter grade} & Effectiveness in & Effectiveness in  \\
          & reducing probability, $\alpha_{ik}$ & reducing consequence, $\beta_{ijk}$ \\
          \hline
          A & 0.90 & 0.80 \\
          B & 0.925 & 0.85 \\
          C & 0.95 & 0.90 \\
          D & 0.975 & 0.95 \\
          \hline
    \end{tabular}
\end{table}


\subsection{Results}
We used MindtPy in Python and Gurobi solver to solve the integer program, the resource allocation model, for the state of Iowa. 
Different budget amounts were used ranging from \$100,000 to \$120 million. 
Figure 2 illustrates the relationship between the objective function (expected cost) and budget amounts less than \$5 million. Not surprisingly, as the budget increases, the objective function decreases, indicating that selecting more projects will mitigate more risk to the state of Iowa. If no projects are selected, the objective function equals \$7.62 billion. If all of the projects are selected with a budget of \$120 million, the objective function equals \$4.46 billion. As seen in Figure 2, the objective function decreases at approximately a linear rate for budgets ranging from \$0 to \$900,000, with a \$2918 improvement in the objective function for every \$1 increase in the budget. 
The rate of improvement in the objective function decreases significantly when the budget exceeds \$1 million. Although not shown in Figure 1, the rate of improvement in the objective function decreases even further if the budget is more than \$5 million. For example, the objective function is \$4.54 billion if the budget equals \$5 million, \$4.50 billion if the budget equals \$10 million, and \$4.46 billion if the budget equals \$60 million. 
Although additional resources reduce the risk, diminishing marginal benefits may indicate that some of the projects should not be selected because the additional cost may not be worth the relatively small reduction in risk.

\begin{figure}[H]
    \centering
    \includegraphics[width=0.6\linewidth]{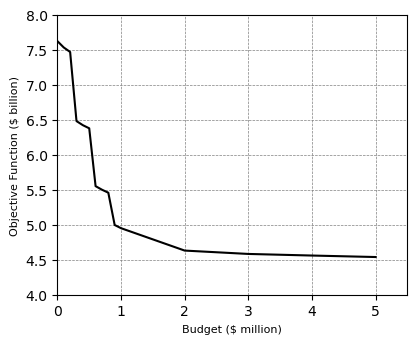}
    \caption{Objective function vs budget}
    \label{fig:ObjectiveFunction}
\end{figure}

Figure 3 presents a detailed analysis of project selection across various budget levels, illustrating the model's decision-making process in optimizing resource allocation. The results consistently indicate that Project 20 is selected across all budget scenarios due to its cost-effectiveness and substantial impact on risk reduction. With a cost of \$24,000, Project 20 is one of the least expensive alternatives and is very effective with a letter grade of A. This project focuses on mitigating floods, addressing a critical hazard for the state of Iowa. Floods have an annual probability of 45\%. Its consistent selection underscores the importance of prioritizing projects that maximize risk reduction benefits at a low cost. 

The model strategically balances costs and benefits, optimizing trade-offs based on budget constraints. For instance, if the budget is  \$600,000, the model recommends Projects 2, 20, 47, and 48. If the budget increases to \$700,000, the selection shifts to Projects 17, 20, 47, 48, and 51. Project 2 is the least expensive alternative, and the combined cost of Projects 17 and 51 are \$126,781 more than Project 2. However, with a letter grade of A, Project 17 is very effective in reducing both the probability and consequences of flooding. Project 51, also graded as an A project, is recommended because of its broad applicability in mitigating the consequences of multiple hazards, including floods, earthquakes, extreme heat, wildfires, hazmat releases, IED attacks, tornadoes, radiological incidents, dam failures, winter storms, and bridge failures. 

The model prioritizes projects that offer the best risk reduction within the given budgetary constraints. Projects with large costs and less effectiveness or projects that address hazards with smaller probabilities or less impact receive less priority. For example, Project 16 is not selected until the budget equals \$120 million—an amount that exceeds the total cost of all 52 projects—indicating that Project 16 is the lowest priority. This is primarily due to its high cost (\$1,787,918) and limited effectiveness, as reflected by its letter grade of D. Additionally, Project 16 is exclusively targeted at wildfire mitigation, a hazard with a relatively low probability of occurrence in Iowa (9.1\%). Similarly, Projects 3 and 35 are only selected if the budget exceeds \$115 million. Project 3, with a cost of \$2,567,719 and a letter grade of B, and Project 35, with a cost of \$3,580,158 with a letter grade of C, both focus solely on flood mitigation. Although flooding is a consequential hazard with a high probability of occurrence, a multitude of other less expensive projects are more effective at mitigating flooding. These findings underscore the model’s capability to systematically allocate resources to projects that yield the highest risk reduction benefits per unit cost, ensuring an optimized balance between affordability and effectiveness.

\begin{figure}[H]
    \centering
    \includegraphics[width=\linewidth]{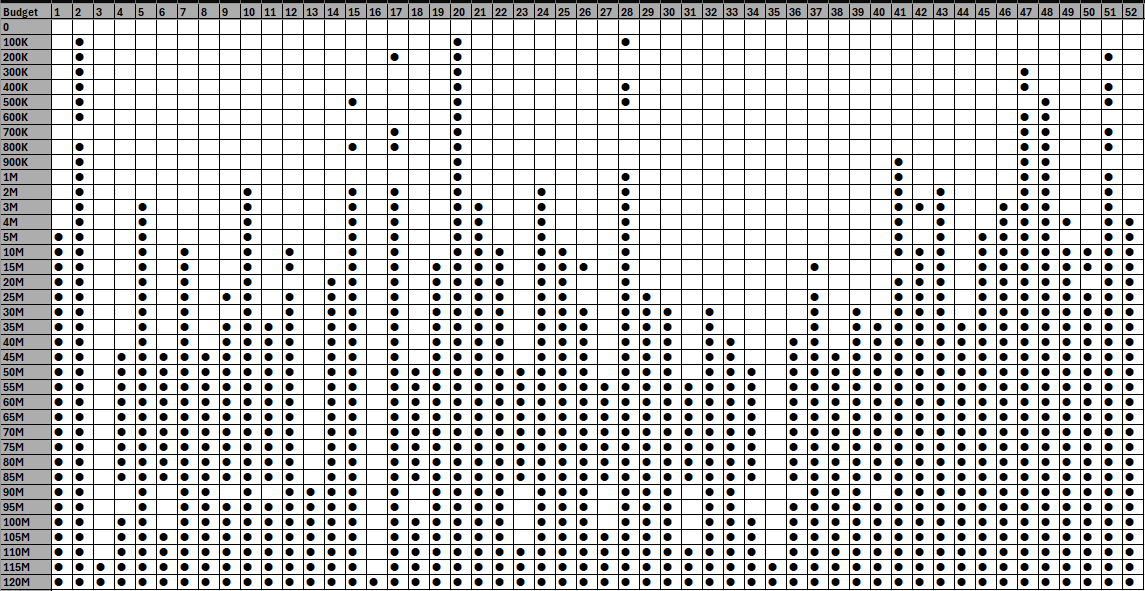}
    \caption{Optimal allocation at different budget amounts}
    \label{fig:ObjectiveFunction}
\end{figure}

\subsubsection{Sensitivity Analysis}
One of the biggest challenges in applying this resource allocation model to the state of Iowa's homeland security and emergency management planning is estimating the numerical values for the effectiveness parameters $\alpha_{ik}$ and $\beta_{ijk}$. The letter grades from the 2023 Iowa Comprehensive Emergency Plan provides a valuable starting point, but we arbitrarily chose how to translate the letter grades to effectiveness numbers as shown in Table \ref{tab:effectiveness}. Sensitivity analysis on the effectiveness numbers was conducted in order to evaluate the extent to which the model's recommendations depend on the specific numbers for effectiveness.
These effectiveness parameters play a crucial role in determining the prioritization of mitigation projects. The parameter $\alpha_{ik}$ assesses a project's ability to reduce the hazard probability and $\beta_{ijk}$ measures the project's ability to reduce the losses from a hazard. Given the uncertainty in estimating these parameters, variations in their values may lead to shifts in resource allocation decisions. 

Three sets of different effectiveness numbers are generated as a new scenario. Each scenario represents a different method of translating the letter grade to the effectiveness value. The effectiveness numbers are presented in Table \ref{tab:scenarios}.  All projects become less effective in the scenario 1. The difference in effectiveness between two consecutive letter grades is 0.01. In the base-case scenario, the difference between two letter grades is 0.025 for $\alpha_{ik}$ and 0.05 for $\beta_{ik}$. A and B projects perform much worse in scenario 1 than in the base-case scenario. 
Scenario 2 tests the effect of assuming that A and B projects are similar in effectiveness at reducing the consequences, but C and D projects are much less effective than A and B projects. Both A and B projects are more effective at reducing the consequences than reducing the probability. The effectiveness of reducing the probability of a hazard for A projects is approximately equal to that of B projects. The effectiveness of C projects is similar to the effectiveness of D projects.   
The results of the model will reveal whether a decision maker should reallocates resources more substantially toward A and B projects or continue to fund C and D projects despite their declining effectiveness. In scenario 3, A, B, and C projects have similar effectiveness values to each other (differences of 0.01 in probability reduction and 0.02 in consequence reduction). D projects' effectiveness is much worse than the other projects' effectiveness. Scenario 3 tests the extent to which allocating resources should select more B and C projects because of their similarity in effectiveness to A projects. 

\begin{table}[H]
    \centering
    \caption{Effectiveness Values for Different Scenarios}
    \label{tab:scenarios}
    \begin{tabular}{c|cc|cc|cc}
        \hline
        \multirow{2}{*}{Letter Grade} & \multicolumn{2}{c|}{Scenario 1} & \multicolumn{2}{c|}{Scenario 2} & \multicolumn{2}{c}{Scenario 3} \\
        & $\alpha_{ik}$ & $\beta_{ijk}$ & $\alpha_{ik}$ & $\beta_{ijk}$ & $\alpha_{ik}$ & $\beta_{ijk}$ \\
        \hline
        A & 0.96 & 0.95 & 0.90 & 0.75 & 0.90 & 0.80 \\
        B & 0.97 & 0.96 & 0.91 & 0.80 & 0.91 & 0.82 \\
        C & 0.98 & 0.97 & 0.98 & 0.95 & 0.92 & 0.84 \\
        D & 0.99 & 0.98 & 0.99 & 0.97 & 0.975 & 0.95 \\
        \hline
    \end{tabular}
    \label{tab:effectiveness_scenarios}
\end{table}

We solved the resource allocation model for these three scenarios for the same budget amounts as in the base-case scenario, from \$100,000 to \$120 million.
The optimal solutions for all three scenarios largely mirror the optimal solutions of the base-case scenario with some changes in project selection at the margin. These results indicate that the specific effectiveness values may not drastically change the project selection as long as the projects have accurate A, B, C, and D letter grades.

To better explore and analyze these scenarios, we discuss the optimal solutions for two budget constraints. Comparing allocations with the different scenarios for different budgets allows us to assess how increasing or decreasing the effectiveness in some projects can change the optimal allocation of projects at the margin. 

If the budget is \$2 million, the project allocations for scenario 1 and 2 are different than in the base-case scenario, but the allocation for scenario 3 is the same as in the base-case. In scenario 1, Projects 21 (A) and 42 (C) replace Project 28 (B). All three projects mitigate flooding, a high-probability hazard. The combined cost of Projects 21 and 42 exceeds that of Project 28. Despite their greater combined cost, the model in scenario 1 favors two projects with similar effectiveness over a single project. The effectiveness parameters in scenario 1 differ by 0.01 whereas they differ by 0.05 in the base-case scenario.

If the budget is \$2 million, Project 24 (D) in the base-case scenario is replaced by Projects 21 (A) and 52 (B) in scenario 2. Project 21 only mitigates flooding, but Projects 24 and 52 mitigate 9 different hazards. Since A and B projects are much more effective at reducing consequences than D projects in scenario 2, the model in scenario 2 recommends the A and B projects.

If the budget is \$20 million, the project allocation for the three scenarios differ more extensively from the base-case allocation. Projects 14 (A) and 19 (C) in the base-case are replaced by ten lower-cost projects in scenario 1: Projects 9 (B), 11 (A), 12 (A), 18 (D), 26 (B), 29 (A), 32 (A), 36 (A), 40 (A), and 50 (D). Most of these projects mitigate flooding, a couple of projects mitigate another hazard, and Project 50 mitigates five hazards. Project 14, which costs \$8.53 million, mitigates both flood and drought, and Project 19, which costs \$3.58 million, mitigates only drought. Since projects are less effective in scenario 1 than in the base-case, the model in scenario 1 prioritizes more projects that mitigate flooding rather than a single, expensive A project that mitigates both flood and drought as in the base-case scenario. 

In scenario 2, Projects 9 (B), 12 (A), 39 (A), and 50 (D) replace Projects 7 (B) and 19 (C) in the base-case scenario with a budget of \$20 million. Project 7 mitigates flooding, and Project 19 mitigates drought. Since A and B projects are more effective in scenario 2, the model recommends cheaper projects with broader impact, prioritizing flood mitigation (Projects 9 and 12) and multi-hazard mitigation (Projects 39 and 50). 

Projects 7 (B) and 22 (A) in the base-case are replaced by Project 37 (C) in scenario 3 with a \$20 million budget. Since C projects are more effective in scenario 3 than in the base-case, the model in scenario 3 recommends Project 37, which mitigates tornadoes (the hazard with the greatest probability) and earthquakes, rather than two projects that only mitigate flooding. Other projects recommended by scenario 3 mitigate flooding.
Similar patterns emerge across other budget amounts for these three scenarios. 
 
Another sensitivity analysis seeks to understand the impact of the consequences on project allocation. Although the probabilities and consequences of the hazards represent extreme events in the resource allocation model, these hazards are less consequential than the scenarios outlined in state of Iowa's THIRA \cite{HSEMD2021}. We change the consequence value for each hazard to the number in the THIRA for fatalities, injuries, customers without power, and businesses closed. These estimates represent quasi-worst-case scenarios for all 16 hazards and allow us to evaluate how the model responds to more extreme conditions. The THIRA does not estimate property losses or crop damage, so those consequences remain the same as in Table \ref{tab:hazards}. 

If the budget is \$2 million, Projects 10 (C), 17 (A), 24 (D), and 43 (B) in the base-case are replaced by Projects 25 (B) and 52 (B). 
Projects 17 and 43 only mitigate flooding. Although Projects 10 and 24 address multiple hazards, their effectiveness is relatively low. The severity of the consequences influences the model to select projects that mitigate very severe risks that are not mitigated by other projects. Project 25 mitigates tornadoes and earthquakes. Tornadoes have a very high probability of occurrence and result in a significant number of fatalities, injuries, customers without power, and business closures in the THIRA. A major earthquake is extremely unlikely in Iowa but it could lead to catastrophic consequences, as reflected in the THIRA. Project 52 is a relatively effective project and mitigates several hazards, floods, cyberattacks, droughts, earthquakes, extreme heat, wildfires, tornadoes, dam failures, and winter storms. 

If the budget is \$20 million, Projects 14 (A) and 19 (C) in the base-case are replaced by eleven less expensive projects: 4 (B), 8 (C), 9 (B), 12 (A), 18 (D), 26 (B), 30 (A), 37 (C), 38 (C), 39 (A), and 50 (D). 
Project 14 mitigates both flood and drought, and Project 19 mitigates drought. 
The model containing more severe consequences recommends more projects that together mitigate more hazards, particularly those hazards with the most severe impacts. For example, Projects 30 and 39, both with a letter grade of A, mitigate multiple hazards: floods, cyberattacks, drought, earthquakes, extreme heat, wildfires, tornadoes, dam failures, and winter storms. This sensitivity analysis, which was conducted for a range of budget amounts from \$100,000 to \$120 million, indicates that including more severe consequences for all hazards should in general result in more projects being selected that mitigate more hazards. 

\section{Conclusion}
This article presents an optimization-based decision support model for allocating resources to mitigate the risk of multiple and varied homeland security risks. 
Given the increasing frequency and severity of disruptive events in the United States, decision makers at the federal, state, and local levels within the homeland security enterprise need to smartly and efficiently allocate limited resources in order to minimize disaster impacts. Our model incorporates multiple hazards, probabilistic risk assessments for each threat and hazard, and multidimensional consequences, enabling a comprehensive approach to selecting projects that minimizes the risk subject to a budget constraint.

Applying this model to the state of Iowa illustrates its usefulness toward guiding resource allocation decisions for HSEMD. The application considers 16 hazards, 6 types of consequences, and 52 mitigation projects for budget amounts ranging from \$100,000 to \$120 million. 
The expected consequences (i.e., the objective function) from these 16 hazards without any mitigation are \$7.6 billion. Most of these losses are due to human disease, which could lead to over 8000 fatalities and 500,000 injuries or hospitalizations (similar to the COVID-19 pandemic). If the human disease scenario is removed from the model, the expected consequences are \$903 million. The National Risk Index calculates the expected annual losses from natural hazards is \$842 million for the state of Iowa \cite{FEMA_NRI2023}. Since the National Risk Index does not include human disease or a pandemic as a natural hazard, the similarities in these two dollar amounts help to validate the probabilistic risk assessments provided in this article. The model in this article contains hazards that are not weather or geological and two more types of consequences than the National Risk Index.

The results indicate that prioritizing highly effective, multi-hazard projects provides the greatest return on investment. Projects with an effectiveness grade of A, especially those that mitigate multiple hazards, are frequently recommended for different budgets. As the budget increases, the expected consequences decreases, but we observe diminishing marginal returns. For example, the first \$1 million of allocation reduces the objective function from \$7.6 billion to \$5.0 billion, but the objective function only decreases to \$4.6 billion if the budget is \$2 million. When the budget exceeds \$30 million, the decrease in the objective function is less than the increase in the budget, which signifies that budgeting more than \$30 million for these homeland security and emergency preparedness projects is not a good use of funds. The risk reduction is not worth the additional cost. If all of the projects are allocated, which requires a budget of \$120 million, the objective function is \$4.5 billion, which is approximately \$100 million less than the objective function with a budget of \$2 million. 

Sensitivity analysis examined the extent to which the assumptions about project effectiveness impact the recommendations of the model. The model recommends many of the same projects in the three sensitivity analysis scenarios as in the base-case scenario. Some differences stand out, however. If projects are less effective than in the base-case, the model recommends selecting more less expensive projects that mitigate multiple hazards. If the consequences are worse than than in the base-case scenario, the model recommends selecting more projects that mitigate a wider range of hazards. These findings suggest that identifying more accurate ways to numerically assess the effectiveness of different projects can provide a better allocation of resources. If the projects' effectiveness follows the letter grade for the projects as presented in this article, then the 70-80\% of the selected projects should follow the base-case recommendations.

While this study provides valuable insights into optimizing risk mitigation investments in homeland security, several avenues for future research remain. First, incorporating stakeholder preferences and political considerations could align resource allocation decisions with broader policy objectives. Second, the model could consider uncertainty in the consequences, the effectiveness parameters, and the cost of each project. Incorporating these uncertain elements will require stochastic programming to identify the optimal resource allocation under uncertainty. Finally, these allocation decisions are typically made on an annual basis as DHS annually provides funds to states for different projects. Modeling decision making over time and how the hazard probabilities may change over time could provide an insightful dynamic decision-making model for homeland security and emergency preparedness.

In conclusion, this research provides a framework for optimizing resource allocation in homeland security and emergency preparedness. By balancing the likelihood and consequences of multiple hazards, risk reduction, cost, and effectiveness, the proposed model serves as a valuable tool for policymakers and emergency management. As natural hazards, human-caused threats, and technological and infrastructure failures continue to pose significant challenges to communities, adopting optimization-based approaches can improve preparedness and resilience, and ultimately save lives and reduce damage.

\newgeometry{left=0.75in, right=0.75in, top=1in, bottom=1in} 
\newpage
\appendix
\section{Appendix}

\renewcommand{\thetable}{A.\arabic{table}}
\setcounter{table}{0}
\small
\begin{longtable}{|>{\centering\arraybackslash}p{3cm}|c|p{8cm}|c|c|}
    \caption{State Mitigation Projects with Costs and Letter Grades \label{sec:appendixa}} \\
    \hline
    \textbf{Category} & \# & \textbf{Project Description} & \textbf{Cost (\$)} & \textbf{Letter Grade} \\
    \hline
    \endfirsthead

    \hline
    \textbf{Category} & $k$ & \textbf{Project Description} & \textbf{Cost (\$)} & \textbf{Letter Grade} \\
    \hline
    \endhead

    \hline
    \multicolumn{5}{r}{\textit{Continued on next page}} \\
    \endfoot
    \endlastfoot
    \multirow{3}{=}{\textbf{Planning and Regulation}}
     & 1 & Provide training or outreach, encouraging implementation of building codes and retrofits & 1,590,427 & A \\
    \cline{2-5}
    & 2 & Facilitate communities to join the Community Rating System. & 16,133 & B \\
    \cline{2-5}
    & 3 & Maintain at least 700 communities in the National Flood Insurance Program. & 2,567,720 & B \\
    \cline{2-5}
    & 4 & Create a guide for communities to manage deed-restricted flood buyout properties & 663,284 & B \\
    \cline{2-5}
    & 5 & Provide training to communities prone to location-specific hazards. & 486,680 & B \\
    \cline{2-5}
    & 6 & Advocate for flood mitigation in watershed plans & 1,687,539 & C \\
    \cline{2-5}
    & 7 & Develop a comprehensive, statewide flood mitigation strategy that considers flood buy-outs, watershed approach flood mitigation, levees, and other solutions & 744,214 & B \\
    \cline{2-5}
     & 8 & Have 100 percent of high hazard potential dams with emergency action plans  & 934,771 & C \\
     \cline{2-5}
    & 9 & Identify public buildings that are in the special flood hazard area and encourage retrofit and mitigation plans & 619,463 & B \\
    \cline{2-5}
     & 10 & Jurisdictions acquire software to help mitigate hazards & 571,229 & C \\
    \cline{2-5}
    & 11 & Ensure that flood impact data and watershed planning initiatives are shared & 1,110,006 & A \\
    \cline{2-5}
    & 12 & Develop at least 3 watershed plans or hydrologic that analyze hazard mitigation options. & 1,000,000 & A \\
    \cline{2-5}
     & 13 & Develop a state-wide electric resiliency strategy & 53,862,450 & A \\
    \cline{2-5}
    & 14 & Encourage local jurisdictions to participate in watershed management authorities recommend mitigation solutions for levee and flood issues and drought & 8,525,723 & A \\
    \cline{2-5}
     & 15 & Develop, implement, and improve the Iowa Drought Plan's communication plan & 72,932 & B \\
    \cline{2-5}
    & 16 & Communities develop a Community Wildfire Protection Plan & 1,787,919 & D \\
    \cline{2-5}
    & 17 & Provide technical assistance to help 15 communities understand their flood issues to explore mitigation alternatives & 80,000 & A \\
    \cline{2-5} 
    \hline
    \multirow{3}{=}{\textbf{Resilient Systems and Structures}} 
    & 18 & Provide dry hydrants in wildland-urban interface areas with no water mains & 115,270 & D \\
    \cline{2-5}
    & 19 & Connect drought-vulnerable water supply systems to other water supplies & 3,580,158 & C \\
    \cline{2-5}
    & 20 & Elevate or protect wastewater lift stations, and/or complete other sanitary sewer hazard mitigation improvements & 24,000 & A \\
    \cline{2-5}
    & 21 & Mitigate flooding of buildings by elevating buildings, flood-proofing, constructing non-levee embankments, or acquiring and removing buildings & 89,994 & A \\
    \cline{2-5}
    & 22 & Increase floodwater storage through floodplain or streambank restoration projects & 344,451 & A \\
    \cline{2-5}
    & 23 & Put in impervious manholes, pumps, or backflow prevention, or similar small-scale flood protection projects & 2,605,091 & B \\
    \cline{2-5}
    & 24 & Install and maintain protective measures for the physical safety and security of critical facilities & 191,067 & D \\
    \cline{2-5}
    & 25 & Construct public safe rooms & 805,598 & B \\
    \cline{2-5}
    & 26 & Reduce damage from flooding and erosion through stream channel improvement projects & 555,262 & B \\
    \cline{2-5}
    & 27 & Rehabilitate dams and levees of high hazard potential & 2,523,207 & C \\
    \cline{2-5}
    & 28 & Provide information to owners of underground storage tanks about potential damages from flooding & 38,12 & B \\
    \cline{2-5}
     & 29 & Implement green infrastructure in cities to mitigate flooding  & 1,403,713 & A \\
    \cline{2-5}
    & 30 & Provide more resilient electric service through: robustness measures, installation of cold weather protection measures, extreme heat and drought resistance measures, flood protection measures, or wind protection measures & 1,727,926 & A \\
    \cline{2-5}
    & 31 & Mitigate flood damage to structures or public facilities to retrofit bridges, elevate roads, or build or reconstruct levees & 2,523,207 & B \\
    \cline{2-5}
    & 32 & Mitigate flooding with a watershed approach by putting in practices that detain water and/or increase infiltration. 
    & 1,403,713 & A \\
    \cline{2-5}
     & 33 & Initiate projects to reduce landslide damage and risk where landslides are most likely & 2,523,207 & D \\
    \cline{2-5}
    & 34 & Reduce water losses through leak detection and/or distribution system renovation projects & 3,580,158 & B \\
    \cline{2-5}
    & 35 & Encourage development of gray water infrastructure, recycling and reusing water & 3,580,158 & C \\
    \cline{2-5}
    & 36 & Develop additional water storage, especially floodwater diversion and storage options & 2,201,108 & A \\
    \cline{2-5}
    & 37 & Encourage the building of shelters (other than safe rooms) at parks and other outdoor areas where people congregate & 991,476 & C \\
    \cline{2-5}
    & 38 & Equip public facilities, community centers, and resilience hubs to act as cooling and warming centers during extreme temperature events & 1,524,312 & C \\
    \cline{2-5}
    & 39 & Install transfer switches, panels, and connections for easy or automatic use of microgrids or generators to supply power & 1,727,926 & A \\
    \cline{2-5}
    & 40 & Encourage programs for residential properties that implement on-site stormwater management practices & 1,403,713 & A \\
    \cline{2-5}
    \hline
    \multirow{3}{=}{\textbf{Cooperation and Awareness}} 
    & 41 & Improve awareness of hazard risks and ways to reduce their impacts through signage projects or awareness campaigns & 306,412 & A \\
    \cline{2-5}
    & 42 & Provide training or outreach to 5 communities with repetitive loss properties, including information about reducing future damage & 43,285 & C \\
    \cline{2-5}
    & 43 & Discuss establishing a state levee safety program and consolidating levee districts at the USACE system level and garner support amongst stakeholders & 100,000 & B \\
    \cline{2-5}
    & 44 & Provide example standards and guides to local jurisdictions that promote green infrastructure practices and measures that direct water away from structures. & 1,094,720 & B \\
    \cline{2-5}
    & 45 & Develop coordinated, prompt, reliable, and accessible information for the whole community concerning current and likely drought and water supply status, drought vulnerability, drought-time response actions, and continuous conservation measures & 358,160 & C \\
    \cline{2-5}
    & 46 & Maintain and expand monitoring network for stream flows, precipitation, soil moisture, evapotranspiration, and groundwater levels, in order to characterize Iowa’s surface and groundwater resource availability & 338,735 & C \\
    \cline{2-5}
    & 47 & Explore the creation of tools that can help communities understand mitigation measures & 254,665 & B \\
    \cline{2-5}
    & 48 & Annually provide training and/or outreach about mitigation opportunities, available resources, and application specifics with a special focus on smaller communities and underserved communities & 276,807 & A \\
    \cline{2-5}
    & 49 & Discuss flood and drought mitigation opportunities with the Iowa Water Resources Coordination Council and Iowa Watershed Planning Advisory Council & 975,345 & B \\
    \cline{2-5}
    & 50 & Encourage water utilities to review their operating procedures to ensure availability of backup or secondary water systems & 921,588 & D \\
    \cline{2-5}
    \hline
    \multirow{4}{=}{\textbf{Warning and Redundancy to Mitigate Disaster Disruptions}} 
    & 51 & Establish or improve warning and alert systems (e.g. sirens) & 62,914 & A \\
    \cline{2-5}
    & 52 &  Provide back-up power generation, storage, or other energy redundancy measures to serve critical facilities or lifelines & 105,403 & B\\
    &&&& \\
    \cline{1-5}
\end{longtable}
\restoregeometry

\bibliographystyle{unsrt} 
\bibliography{library} 

\end{document}